\documentclass[aps,prl,twocolumn,superscriptaddress,showpacs]{revtex4-1}

\usepackage{amssymb,amsfonts,amsmath}
\usepackage{units} 
\usepackage{braket}
\usepackage{relsize} 

\usepackage{epsfig}
\usepackage{graphicx}

\makeatletter
\DeclareRobustCommand{\genericinterval}[2]{%
  \@ifstar{\genericinterval@star{#1}{#2}}{\genericinterval@nostar{#1}{#2}}}
\newcommand{\genericinterval@star}[4]{\mathopen{}\mathclose{\left#1#3,#4\right#2}}
\newcommand{\genericinterval@nostar}[4]{\mathopen{#1}#3,#4\mathclose{#2}}

\makeatother

\begin{document}
\newcommand{\vq}{\mathbf{q}}
\newcommand{\vp}{\mathbf{p}}
\newcommand{\vP}{\mathbf{P}}
\newcommand{\Wcm}{~Wcm$^{-2}\,$}
\newcommand{\Ea}{E_\mathrm{a}}
\newcommand{\Up}{U_\mathrm{p}}
\newcommand{\Upm}{U_{\mathrm{p},\max}}
\newcommand{\tp}{\tau_\mathrm{t}}
\newcommand{\tr}{\tau_\mathrm{r}}
\newcommand{\Tp}{T_\mathrm{p}}
\newcommand{\tg}{\tau_\mathrm{g}}
\newcommand{\vE}{\mathbf{E}}
\newcommand{\vA}{\mathbf{A}}
\newcommand{\ver}{\mathbf{r}}
\newcommand{\valpha}{\mbox{\boldmath{$\alpha$}}}
\newcommand{\vpi}{\mbox{\boldmath{$\pi$}}}
\newcommand{\Rep}{\mathrm{Re}\,}
\newcommand{\Imp}{\mathrm{Im}\,}
\newcommand{\AL}{A_\mathrm{L}}
\newcommand{\ve}{\hat{\mathbf{e}}}
\newcommand{\Ep}{E_\mathbf{p}}

\newcommand{\etal}{\emph{et al.~}}

\title{Steering proton migration in hydrocarbons using intense few-cycle laser fields}

\author{M. K\"ubel}
\affiliation{Department of Physics, Ludwig-Maximilians-Universit\"at Munich, D-85748 Garching, Germany}
\author{R. Siemering}
\affiliation{Department of Chemistry and Biochemistry, Ludwig-Maximilians-Universit\"at Munich, D-81377 M\"unchen, Germany}
\author{C. Burger}
\affiliation{Department of Physics, Ludwig-Maximilians-Universit\"at Munich, D-85748 Garching, Germany}
\author{Nora G. Kling}
\affiliation{Department of Physics, Ludwig-Maximilians-Universit\"at Munich, D-85748 Garching, Germany}
\affiliation{J.R. Macdonald Laboratory, Physics Department, Kansas-State University, Manhattan, KS66506, USA}
\author{H. Li}
\affiliation{Department of Physics, Ludwig-Maximilians-Universit\"at Munich, D-85748 Garching, Germany}
\affiliation{Max Planck Institute of Quantum Optics, D-85748 Garching, Germany}
\author{A.S. Alnaser}
\affiliation{Max Planck Institute of Quantum Optics, D-85748 Garching, Germany}
\affiliation{Physics Department, American University of Sharjah, POB26666, Sharjah, UAE}
\author{B. Bergues}
\affiliation{Max Planck Institute of Quantum Optics, D-85748 Garching, Germany}
\author{S. Zherebtsov}
\affiliation{Department of Physics, Ludwig-Maximilians-Universit\"at Munich, D-85748 Garching, Germany}
\affiliation{Max Planck Institute of Quantum Optics, D-85748 Garching, Germany}
\author{A. M. Azzeer}
\affiliation{Department of Physics \& Astronomy, King-Saud University, Riyadh 11451, Saudi Arabia}
\author{I. Ben-Itzhak}
\affiliation{J.R. Macdonald Laboratory, Physics Department, Kansas-State University, Manhattan, KS66506, USA}
\author{R. Moshammer}
\affiliation{Max Planck Institute of Nuclear Physics, D-69117 Heidelberg, Germany}
\author{R. de Vivie-Riedle}
\affiliation{Department of Chemistry and Biochemistry, Ludwig-Maximilians-Universit\"at Munich, D-81377 M\"unchen, Germany}
\author{M.F. Kling}
\affiliation{Department of Physics, Ludwig-Maximilians-Universit\"at Munich, D-85748 Garching, Germany}
\affiliation{Max Planck Institute of Quantum Optics, D-85748 Garching, Germany}

\date{February 23, 2016}

\begin{abstract}
{Proton migration is a ubiquitous process in chemical reactions related to biology, combustion, and catalysis. Thus, the ability to manipulate the movement of nuclei with tailored light, within a hydrocarbon molecule holds promise for far-reaching applications.  Here, we demonstrate the steering of hydrogen migration in simple hydrocarbons, namely acetylene and allene, using waveform-controlled, few-cycle laser pulses. The rearrangement dynamics are monitored using coincident 3D momentum imaging spectroscopy, and described with a widely applicable quantum-dynamical model. Our observations reveal that the underlying control mechanism is due to the manipulation of the phases in a vibrational wavepacket by the intense off-resonant laser field.}
\end{abstract}

\maketitle

The rearrangement of hydrocarbon bonds via the migration of a hydrogen atom can result in major deformations of molecular architecture, and thus alter the molecule`s chemical properties. 
Examples include keto-enol tautomerism where the migration of a proton changes an aldehyde into an alcohol.
Isomerization reactions of that kind have been the subject of numerous studies \cite{Levine2007,Bandara2012}. Of particular interest was to determine the so-called isomerization time, which has been measured to be within several tens of femtoseconds in small hydrocarbons \cite{Osipov2003, Xu2009, Jiang2010, Ibrahim2014}. 
The phenomenon has also been observed in larger molecules, such as protonated triglycine \cite{Rodriquez2001}. Tracing of the hydrogen migration from different locations within the molecule has been made possible via isotope labeling, see e.g. \cite{Okino2012, Okino2012a, Heazlewood2011}.
The ability to exert control over the migration could lead to advancement in topics such as the efficiency of catalytic reactions \cite{Pines1981} and combustion reactions regarding fuel and energy research \cite{AshtonActon2012}. Furthermore, light-induced control of hydrogen migration may open new reaction pathways which cannot materialize by other means. 

Despite its direct relevance to applied chemistry, studies regarding the control of the hydrogen migration process have been scarce, and have been limited to theory \cite{Uiberacker2004,Madjet2013} for a long time. However, recent progress has been made in coherently controlling isomerization reactions using fundamental parameters of ultrafast strong-field laser sources. Xie, \etal varied the pulse duration and intensity to explore the isomerization of ethylene \cite{Xie2014}, and reported control of the total fragmentation yields of various hydrocarbons \cite{Xie2012}.

Here, we demonstrate steering of the direction of the hydrogen migration using the electric-field waveform of intense few-cycle laser pulses. This approach goes beyond earlier work on toluene \cite{Kaziannis2014} and methanol \cite{Kotsina2015} using two-color pulses with a duration of tens of fs. In contrast, the duration of our few-cycle laser pulses is significantly shorter than the time scale of the isomerization dynamics, therefore avoiding charge-resonance-enhanced ionization \cite{Zuo1995} occurring at large internuclear distances \cite{Kling2006a}. Moreover, an influence of electron localization-assisted enhanced ionization on the dissociation reactions, recently demonstrated for acetylene theoretically \cite{Lotstedt2012,Lotstedt2013} and experimentally \cite{Gong2014}, is also avoided. In comparison to earlier experiments using longer pulses, our experiment therefore allows the study of the pure predetermination of a molecular reaction, in the sense that the laser field initiates a reaction but is not present during the actual reaction.

The electric field of a few-cycle laser pulse can be described in the time domain as $E(t) = E_0(t) \cos{\left(\omega t + \phi\right)}$, where $E_0(t)$ is the envelope, $\omega$ the carrier frequency, and $\phi$ is the phase between the carrier and the envelope (carrier envelope phase; CEP). The CEP determines the electric field waveform on a sub-femtosecond timescale and has proven to be a powerful parameter for controlling electron motion in various systems \cite{Paulus2001, Baltuska2003, Kling2006a, Schiffrin2013}. It has also recently been shown to apply to the control of nuclear wavepackets \cite{Alnaser2014}.

In the present work, we show that the CEP of a few-cycle laser pulse further permits to steer complex photochemical reactions involving structural rearrangements towards a desired outcome. This goes beyond steering the directional ion emission from a molecule undergoing a simple dissociation reaction, such as deprotonation \cite{Miura2014,Alnaser2014}. Using a reaction microscope (REMI) \cite{Ullrich2003}, we observe asymmetric emission of carbon ions from acetylene and trihydrogen ions from allene as a function of the CEP of the driving laser pulses. The results are interpreted in terms of a quantum mechanical model where the direction of hydrogen migration results from the phases of a superposition of vibrational modes, manipulated by the laser CEP.

The experimental setup will be described in detail elsewhere. Briefly, 
hydrogen migration in acetylene and allene is induced and controlled by 
intense, CEP-stable 4-fs laser pulses at a carrier wavelength of 750\,nm 
and 10\,kHz repetition rate. The near-single-cycle pulses are obtained 
from the SMILE laser system at the Ludwig-Maximilians-Universit\"at (LMU), which is based on the Femtolasers 
Femtopower HR CEP4 amplified laser system. The CEP of the laser is stabilized using the feed-forward technique \cite{Lucking2012} and measured using either a stereo-ATI phase meter \cite{Rathje2012} (in the acetylene experiments) or an f-2f interferometer \cite{Schultze2011} (in the allene experiments).

In the REMI, the linearly polarized laser pulses are focused ($f=\unit[17.5]{cm}$) into a cold gas jet of neutral hydrocarbon molecules.  
A homogeneous electric field is used to direct all ions generated in the laser focus onto a time- and position-sensitive multi-hit capable detector, providing the three-dimensional (3D) momentum distributions of multiple ions in coincidence.

To study the strong-field steering of hydrogen-bond rearrangements in acetylene (HCCH), we focus on its isomerization to vinylidene (CCH$_2$). This channel necessarily involves the migration of one of the protons from one side of the molecule to the other. Vinylidene formation is clearly identified by the dissociation of the dication into C$^+$ and CH$_2^+$ fragments that obey momentum conservation (neglecting the small momenta of the emitted electrons). The distinct separation between the fragments from the acetylene and vinylidene breakups, which have similar mass and momenta, is challenging and feasible so far only by coincidence detection methods \cite{Ullrich2003, Alnaser2006}. 

\begin{figure}[ht]
\centerline{\includegraphics[width=0.47\textwidth]{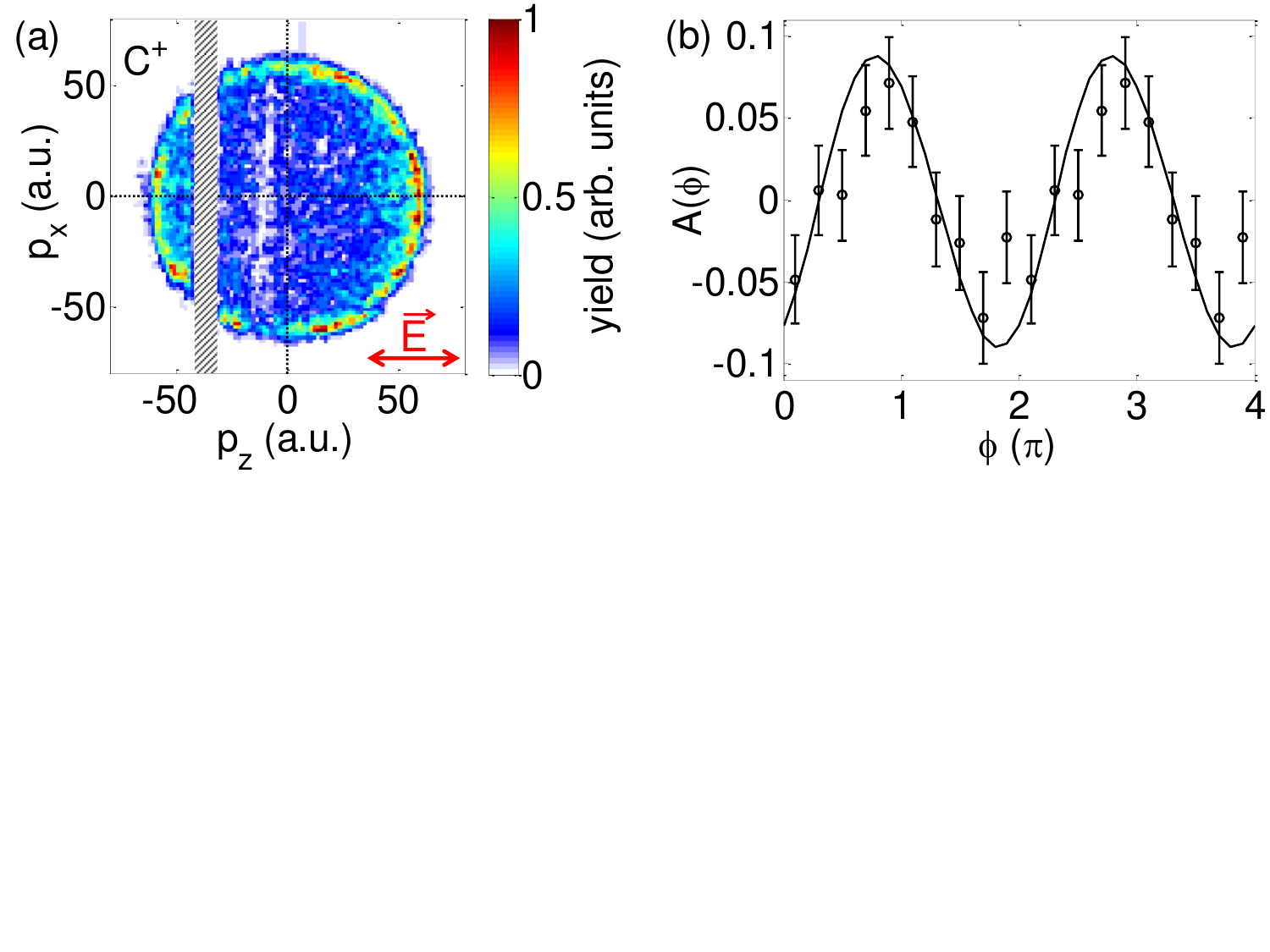}}
\vspace{-3.2cm}
\caption{Strong-field steering of hydrogen migration in acetylene. (a) Measured CEP-averaged momentum distribution for C$^+$ ions emitted from vinylidene dications following hydrogen migration in acetylene, induced by 4-fs laser pulses with an intensity $I \approx 1.3 \times 10^{14}\unit{W/cm^2}$. The shaded area is cut out for technical reasons (see text). The red arrow indicates the axis of laser polarization. (b) CEP-dependent asymmetry parameter for the C$^+$ yield shown in (a), evaluated for a cone of $80^\circ$ around the polarization axis of the laser. Since the absolute value of the CEP is not known from the experiment, the phase is given relative to the asymmetry parameter recorded for C$_2$H$_2^+$ recoil ions (see text). The solid line depicts the calculated asymmetry curve for an intensity of $1.5 \times 10^{14}\,\unit{W/cm^2}$, shifted along the CEP axis for best agreement with the experiment.}
\label{fig:results_acetylene}
\end{figure}

Fig.~\ref{fig:results_acetylene}a shows the momentum along the laser propagation axis ($p_x$) versus the momentum along the laser polarization ($p_z$), for C$^+$ ions produced in the breakup of the vinylidene dication. The nearly isotropic momentum distribution of the C$^+$ ions can be understood  as a consequence of the hydrogen migration preceding dissociation of vinylidene along the C-C bond \cite{Alnaser2006}. For C$^+$ momenta in the indicated range $\unit[-42]{a.u.}<p_z<\unit[-28]{a.u.}$, the C$^+$ ions arrive at the ion detector almost simultaneously with the coincident CH$_2^+$ fragments, which impedes their coincident detection. Hence only events with momenta outside the range $\unit[28]{a.u.}<|p_z|<\unit[42]{a.u.}$ are considered for evaluating  the CEP-dependence of the C$^+$ emission direction. 
The emission direction of the C$^+$ ion following the cleavage of the C-C bond is directly indicative of whether the left proton in acetylene has migrated to the right side (along the laser polarization), or vice versa. The yields of C$^+$ ions emitted to the right, $R(\phi)$, or left, $L(\phi)$, as a function of the CEP are therefore a measure of the of the influence of the CEP on the hydrogen migration. To quantify the CEP dependence, we introduce the asymmetry parameter as 
\begin{equation}
A(\phi)=\left(L(\phi)-R(\phi)\right) / \left(L(\phi)+R(\phi)\right).
\label{eq:A}
\end{equation}
The CEP-dependence of the asymmetry parameter $A_{\mathrm{C}^+}(\phi)$, recorded for C$^+$ ions and plotted in Fig.~\ref{fig:results_acetylene}(b), demonstrates the steering of the direction of proton migration in acetylene. 

For allene (H$_2$C$_3$H$_2$) molecules, the hydrogen motion is monitored through the formation of trihydrogen ions (H$_3^+$). This reaction channel is identified by the dissociation of the allene dication into H$_3^+$ and C$_3$H$^+$ fragments. Even though allene's isomer propyne (HC$_3$H$_3$) is stable in the neutral state and contained in low concentrations in commercially available allene gas bottles, H$_3^+$ formation has been found to result predominantly from the isomerization of allene \cite{Okino2012}.

\begin{figure}[ht]
\centerline{\includegraphics[width=0.47\textwidth]{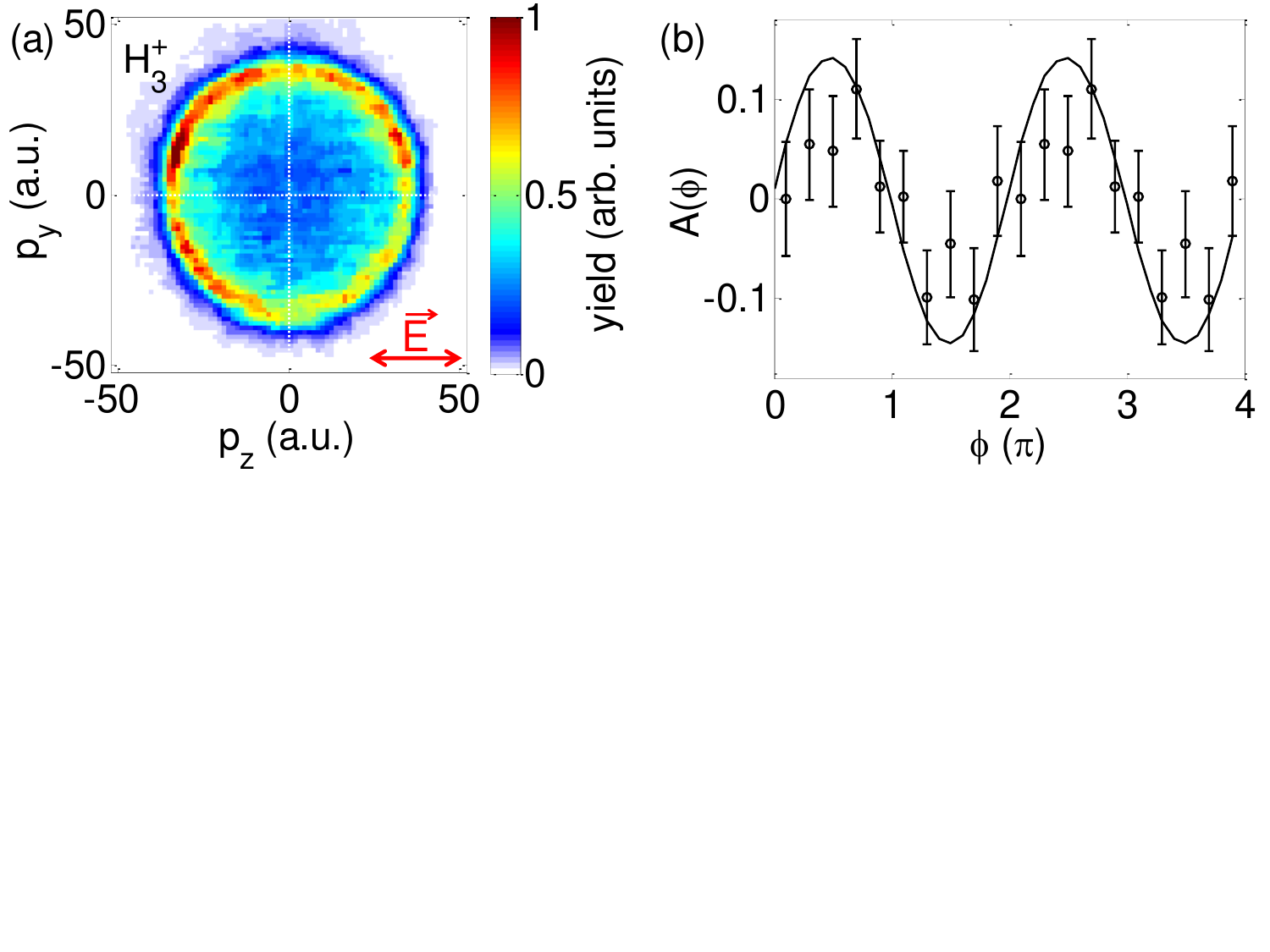}}
\vspace{-3.5cm}
\caption{Strong-field steering of hydrogen migration in allene. (a) CEP-averaged momentum distribution  and (b) CEP-dependent asymmetry parameter (b) for H$_3^+$ ions emitted from propyne dications following hydrogen migration in allene. The red arrow in (a) indicates the axis of laser polarization. The CEP-dependence of the H$_3^+$ is evaluated for a cone of $80^\circ$ around the polarization axis of the laser and at a peak intensity of $\approx 2.9 \times 10^{14}\unit{W/cm^2}$.  The phase is given relative to the asymmetry parameter recorded for C$_3$H$_4^+$ recoil ions (see text). The measured asymmetry curve is compared to the one calculated for an intensity of $3.0 \times 10^{14}\unit{W/cm^2}$, shifted along the CEP axis by the same value as the theory curve in Fig. \ref{fig:results_acetylene}(b).}\label{fig:results_allene}
\end{figure}

The results for the H$_3^+$ formation from allene are displayed in Fig.~\ref{fig:results_allene}. The momentum distribution recorded for H$_3^+$ (Fig.~\ref{fig:results_allene}(a)) is nearly isotropic which is consistent with the findings of Refs \cite{Xu2009, Okino2012}. The CEP dependence of the asymmetry parameter $A_{\mathrm{H}_3^+}(\phi)$ demonstrates the steering of the direction of proton migration in allene. 
Interestingly, we find in our experiments that the asymmetry in the H$_3^+$ formation from allene exhibits a pronounced sensitivity to the laser intensity \cite{Kubel2015}.

The CEP axes in Figs.~\ref{fig:results_acetylene}(b) and \ref{fig:results_allene}(b) are given relative to the phase of the asymmetry parameters (see eq. \ref{eq:A}) measured for the recoil cations of either molecule, produced by single ionization. This makes a comparison of the phases of the asymmetry parameters recorded for hydrogen migration in acetylene and allene possible, indicating that $A_{\mathrm{C}^+}(\phi)$ and $A_{\mathrm{H}_3^+}(\phi)$ assume maxima for similar but not identical CEP values. The experimental data is further compared to predictions from our quantum mechanical calculations. Because the absolute CEP in the experiment is not known, the predicted curve for C$^+$ emission from acetylene was shifted along the CEP for best agreement with the experiment. The prediction for H$_3^+$ emission from allene was shifted by the same amount and lines up with the measured asymmetry. Moreover, the amplitudes of the calculated asymmetries agree reasonably well with the measured ones, lending support to the model described in the following.

\begin{figure}[ht]
\includegraphics[width=.47\textwidth]{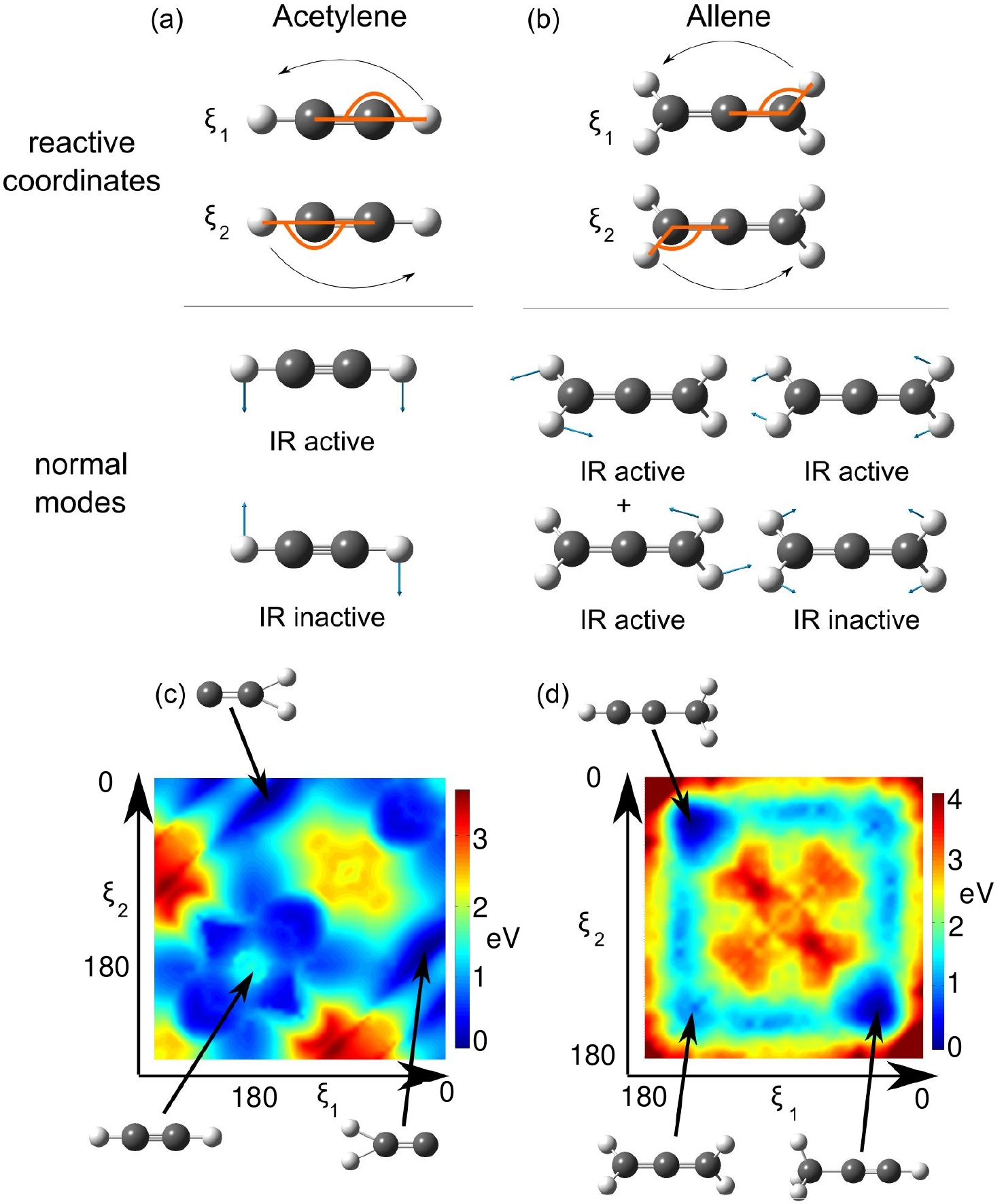}
\caption{Reactive coordinates and normal modes used in the theoretical description of the hydrogen migration reactions in acetylene and allene. The wavepacket formation in acetylene (allene) is calculated on two- (three-) dimensional potential energy surfaces (PES) along the normal modes shown in panels (a)((b)). The reactive coordinates $ \left(\xi_1, \xi_2\right)$.  $ \left(\xi_1, \xi_2\right)$ represent the CCH bond angles (shown in orange) and are used to describe the wavepacket propagation leading to isomerization by hydrogen migration from the left to the right, or from the right to the left. (c) The PES of the A$^3\Pi$ state of the acetylene dication, on which isomerization occurs. The binding energy is plotted as a function of the two reactive coordinates. The initial form of the molecule and the two isomerized configurations are indicated. (d) Same as (c) for the B$^3\Pi$ state of the allene dication, relevant for the isomerization reaction which leads to H$_3^+$ formation.}\label{fig:theory}
\end{figure}

In general any nuclear motion, can be expressed in the basis of all normal modes. We use normal modes to describe the molecule laser interaction and reactive coordinates for the description of the subsequent hydrogen migration. The subset of normal modes relevant for the initiation of the reactions are illustrated in Fig.~\ref{fig:theory}. 
For acetylene (Fig.~\ref{fig:theory}(a)) the IR-active cis-bending mode $\ket{n0}$ and the IR-inactive trans-bending mode $\ket{0m}$ form a 2D basis $\ket{nm}$, where $m, n$ are the number of vibrational quanta. Here, the respective time evolution factor $\exp{\left(-i \frac{E_{m/n}}{\hbar} t\right)}$ is implicitly included and the relative phase of the eigenfunctions is set to compensate any phase offset relative to the CEP. For allene (Fig.~\ref{fig:theory}(b)) three normal modes (IR-active rocking and anti-symmetric bending, and IR-inactive symmetric bending) are required to describe the initiation of isomerization by the laser field.

The interaction of the molecules with the external laser field is calculated by solving the time-dependent Schr\"odinger equation (TDSE) using the Chebychev propagator. The TDSE for the neutral molecule under the influence of the light field reads:
\begin{equation}
\left(H_n + \mu_{nn} \epsilon \left( t \right) \right) \Psi_n \left( t \right) = i \hbar \dfrac{d}{dt} \Psi_n (t),
\end{equation}
where $ H_n$ is the Hamiltonian of $ X^1\Sigma_g^+$, $ \mu_{nn}$ is the associated dipole moment, and $ \Psi_n \left( t \right)$ is the nuclear wave function. The light-field $ \epsilon(t)$ is included in the dipole approximation and is characterized by a full width at half maximum (FWHM) of the intensity envelope of 4\,fs. 

The excitation/ionization mechanism is explained exemplarily for acetylene. The laser only addresses the IR-active modes $\ket{n 0}$, in which it creates a wavepacket $e^{-i \phi} \ket{n 0}$. This wavepacket forms temporarily even if the laser pulse is non-resonant to the vibration, and the phase $\phi$ of the vibrational wavepacket matches the laser CEP.
For our simulations, we consider that the molecule is ionized at the peak intensity of the laser pulse. This approximation leads to a slight overestimation of the asymmetry amplitude, which is consistent with Figs \ref{fig:results_acetylene}(b) and \ref{fig:results_allene}(b). Without the ionization, the molecule would remain in the vibrational ground state after the laser pulse has passed.

As the eigenfunctions of the molecular cation are slightly different from the ones of the neutral molecule, the ionization leads to population of several cationic modes, including the IR-inactive ones  $\ket{0 m}$.
This population process by ionization is independent of the CEP. 
The CEP-dependent part of the full wavepacket is therefore 
predominantly given by the basic superposition of the fundamental IR-active and the IR-inactive modes, 
\begin{equation}
\Psi_\mathrm{basic} = \ket{0 1} + e^{-i \phi} \ket{1 0}. 
\end{equation} 
For interpretation of the control mechanism, it is sufficient to concentrate on $\Psi_\mathrm{basic}$, because it is responsible for the asymmetric isomerization occurring in the dication. The dication is populated 3/4 of an optical cycle after the first ionization, which allows enough time for electron recollision \cite{Niikura2002}. The (full) wavepacket is projected onto the reactive, excited dicationic state, which is the lowest excited state we find to support isomerization.

In the second simulation step, the isomerization dynamics in the reactive state are described on two dimensional PES along two reactive coordinates illustrated in Fig.~\ref{fig:theory}. These coordinates describe the CCH bond angles leading to isomerization. To calculate the isomerization we project the wavepacket $\Psi_\mathrm{modes}$, expressed in normal modes, onto the wavepacket $\psi_\mathrm{reac}$, expressed in reactive coordinates. The projection is performed via the respective eigenfunctions of the dication ground state 

\begin{equation}
\psi_\mathrm{reac} = \sum_{i,j} \Braket{\Psi_\mathrm{modes} | \Phi_i} \Braket{\Phi_i | \phi_j} \ket{\phi_j},
\end{equation}
where $\Phi_i$ are the normal mode eigenfunctions and $\phi_j$ are the reactive coordinate eigenfunctions. 

The isomerization is simulated through evolution of the prepared wavepacket on the reactive states for 480\,fs with time intervals of 0.24\,fs. The reactive states are the $A^3\Pi$ state of the acetylene dication (Fig.~\ref{fig:theory}(c)) and the $B^3\Pi$ state of the allene dication (Fig.~\ref{fig:theory}(d)). On both reactive states the Franck-Condon region is not the global minimum. Isomerization occurs if small barriers in the PES are overcome by the kinetic energy of the generated wavepackets and the isomerization minima indicated in Figs \ref{fig:theory}(c) and (d), are reached. From these isomerization minima, dissociation can take place. The dissociation yield $L$ ($R$) corresponding to the left (right) isomerization is quantified by the part of the wavepacket at the corresponding isomerization minimum after integration over the propagation time. The asymmetry parameter $ A$ is calculated from $L$ and $R$ according to Eq.~\ref{eq:A}.

\begin{figure}[ht]
\centerline{\includegraphics[width=.47\textwidth]{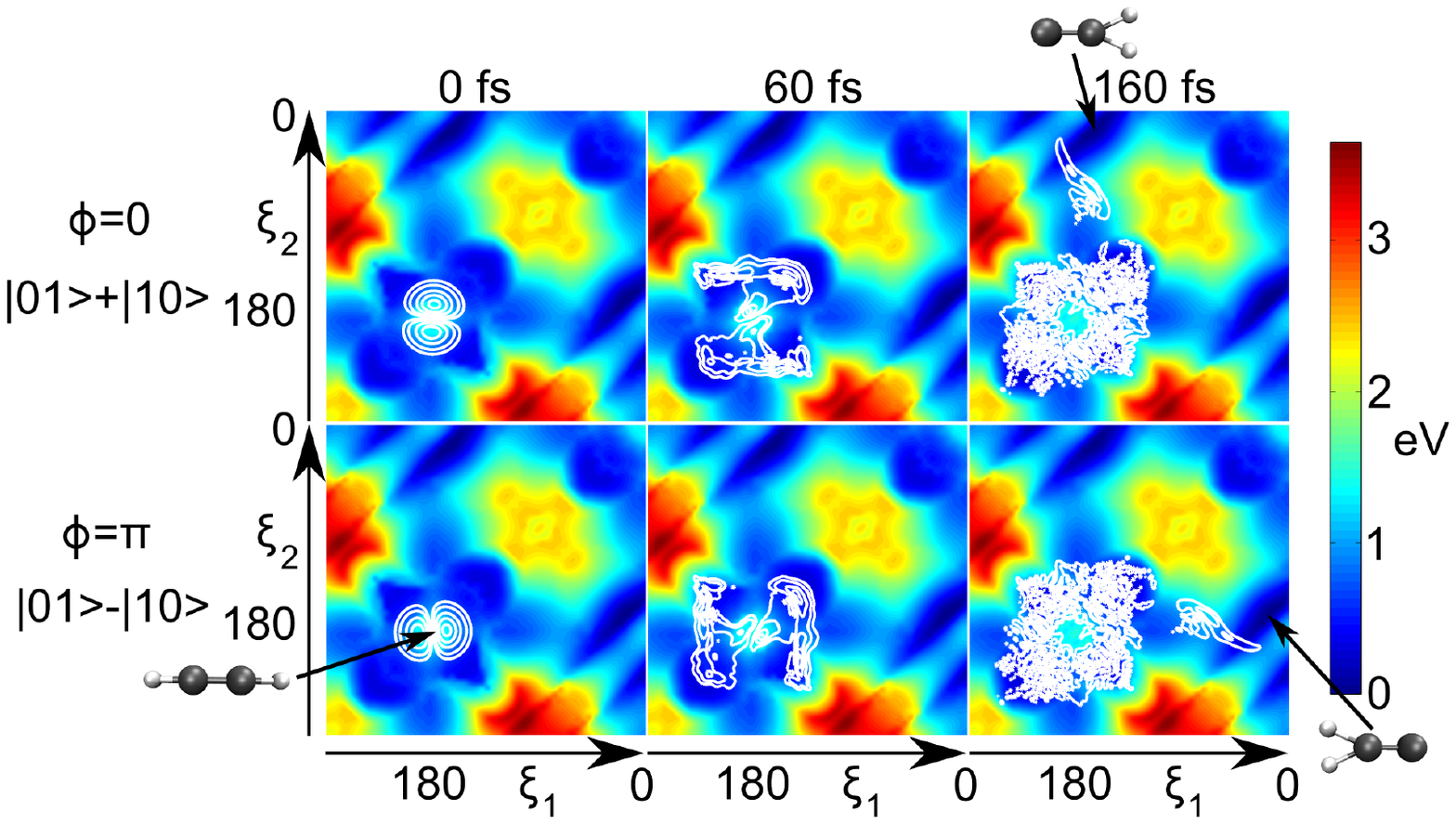}}
\vspace{-1cm}
\caption{Schematic of the control mechanism for hydrogen migration in acetylene. Vibrational wavepackets of the $\Psi_\mathrm{basic}(0)= \ket{01}+\ket{10}$ (top row) and the $\Psi_\mathrm{basic}(\pi) = \ket{01}-\ket{10}$ (bottom row) superposition (see text), are propagated on the PES of the first excited state of the acetylene dication along two reactive coordinates. The contour lines display snapshots of the calculated nuclear wavepacket which leads to isomerization of the acetylene dication. The CEP of the few-cycle pulse determines the sign in the superposition and thereby influences whether the left proton moves to the right (top row) or the right proton to the left (bottom row).}\label{fig:wavepacket}
\end{figure}

For the isomerization of acetylene to vinylidene, the control mechanism is illustrated by the calculated wavepacket propagation displayed in Fig.~\ref{fig:wavepacket}. The wavepacket, that is prepared in the neutral and cation, is projected onto the reactive $A^3\Pi$ state in the dication and propagated along the two reactive coordinates introduced above. The wavepacket possesses a CEP-dependent preferential propagation direction, contained in the basic wavepacket $\Psi_\mathrm{basic}$. The propagation direction results in population of either one of the two indicated potential minima corresponding to the left or the right vinylidene configurations. This leads to the experimentally observed bias in the C$^+$ emission direction after dissociation of the C-C bond, tunable with the CEP.

The isomerization of allene to propyne occurs on the PES plotted in Fig.~\ref{fig:theory}(d) in a similar manner as the isomerization of acetylene but is more complex. The reactive coordinates for the allene isomerization are a linear combination of eight normal modes. Three of these modes (one IR-inactive, two IR-active ones, see Fig.~\ref{fig:theory}(b)) are populated in the preparation step, giving rise to a CEP-dependent wavepacket $\Psi = \ket{0 0 m} + \exp^{-i \phi} (\ket{n_1 0 0}+\ket{0 n_2 0})$, which allows for steering of the direction of hydrogen migration. We believe that the proposed model can be extended to dynamics in even larger molecules -- including rearrangements of functional groups, such as the keto-enol tautomerization -- as long as the proper reaction coordinates can be identified.

In conclusion, we have demonstrated experimentally that few-cycle laser pulses can be exploited to steer the preferential direction of hydrogen migration in small hydrocarbon molecules. The hydrogen migration processes leading to the isomerization of acetylene to vinylidene and to H$_3^+$ formation in allene can be understood in terms of a nuclear control mechanism, similar to the one proposed in Ref.~\cite{Alnaser2014}. According to this mechanism, a field-sensitive nuclear wavepacket is prepared during the few-femtosecond interaction of the molecules with the laser field which predetermines the outcome of an isomerization reaction taking place on a significantly longer time scale. 
The efficiency of the control could be improved by resonant excitation of the vibrational modes and driving ionization by even shorter laser pulses. 
Our work has implications on the strong-field coherent control of more complex photochemical reactions which may lead to alteration of the the structural and functional properties of hydrocarbon molecules.

\begin{acknowledgments}
We are grateful to F. Krausz for his support and fruitful discussions. We acknowledge A. Kessel and S.A. Trushin for experimental support. We are grateful for support by the LMU through the LMUexcellent program, by the Max Planck Society and by the DFG via the Cluster of Excellence: Munich Center for Advanced Photonics (MAP). A.S.A acknowledges support from the American University of Sharjah and the Arab Fund for Economic and Social Development (State of Kuwait). I.B.I was supported by the Chemical Sciences, Geosciences, and Biosciences Division, Office of Basic Energy Sciences, Office of Science, U.S. Department of Energy. We are also grateful for support from the King-Saud University in the framework of the MPQ-KSU-LMU collaboration.
\end{acknowledgments}

\bibliography{isomerization}

\end{document}